\documentclass[11pt]{article}

\usepackage{a4wide}
\usepackage{graphicx}
\usepackage{amsmath}
\usepackage{cite}
\usepackage{authblk}

\usepackage[colorlinks=true,linkcolor=blue,citecolor=blue]{hyperref}

\def\comment#1{}
\def\email#1{}

\def\z{\mathbf{z}}
\def\G{\mathcal{G}}
\def\N{\mathcal{N}}
\def\la{\langle}
\def\ra{\rangle}

\newcommand\GG[1]{\G_{#1}}

\title{\textbf{Multi-loop Integrand Reduction\\ via Multivariate Polynomial Division}\footnote{Presented at the conferences: EPSHEP 2013, Matter to the Deepest 2013, and RADCOR 2013}
}

\date{}

\comment{
} 

\author[1]{Hans van Deurzen}
\author[1]{Gionata Luisoni}
\author[1,2]{Pierpaolo Mastrolia}
\author[1]{Edoardo Mirabella}
\author[3,4]{Giovanni Ossola}
\author[1]{Tiziano Peraro\thanks{Speaker}}  
\author[1]{Ulrich~Schubert}
\affil[1]{\emph{\small Max-Planck Insitut f\"ur Physik, M\"unchen, Germany}}
\affil[2]{\emph{\small Dipartimento di Fisica e Astronomia, Universit\`a di Padova, and INFN Sezione di Padova, Italy}}
\affil[3]{\emph{\small Physics Department, New York City College of Technology, The City University of New York, USA}}
\affil[4]{\emph{\small The Graduate School and University Center, The City University of New York, USA}}

\begin{document}

\maketitle

      \begin{abstract}
        We present recent developments on the topic of the
        integrand reduction of scattering amplitudes.  Integrand-level
        methods allow to express an amplitude as a linear combination
        of Master Integrals, by performing operations on the
        corresponding integrands.  This approach has already been
        successfully applied and automated at one loop, and recently
        extended to higher loops.  We describe a coherent framework
        based on simple concepts of algebraic geometry, such as
        multivariate polynomial division, which can be used in order
        to obtain the integrand decomposition of any amplitude at any
        loop order.  In the one-loop case, we discuss an improved
        reduction algorithm, based on the application of the Laurent
        series expansion to the integrands, which has been implemented
        in the semi-numerical library {\sc Ninja}.  At two loops, we
        present the reduction of five-point amplitudes in
        $\mathcal{N}=4$ SYM, with a unitarity-based construction of
        the integrand.  We also describe the multi-loop
        divide-and-conquer approach, which can always be used to find
        the integrand decomposition of any Feynman graph, regardless
        of the form and the complexity of the integrand, with purely
        algebraic operations.        
      \end{abstract}

\clearpage

\section{Introduction}
Scattering amplitudes are analytic functions of the kinematic
variables.  The investigation of the multi-particle factorization
properties fulfilled at their singular points
\cite{Cachazo:2004kj,Britto:2004ap,Britto:2005fq,Bern:1994zx,Britto:2004nc},
lead to important achievements in the development of methods for their
computation.  Integrand reduction methods, developed for one-loop
diagrams~\cite{Ossola:2006us,Ellis:2007br} and recently extended to
higher
loops~\cite{Mastrolia:2011pr,Badger:2012dp,Zhang:2012ce,Mastrolia:2012an,Mastrolia:2013kca},
allow to express scattering amplitudes as linear combinations of
Master Integrals (MI's), by exploiting the knowledge of the analytic
and algebraic structure of their integrands.

The integrand of a Feynman diagram is a rational function of the
components of the loop momenta, namely a polynomial numerator sitting
over a set of quadratic loop denominators corresponding to internal
propagators.  Any integrand can be written as a combination of
fundamental, \emph{irreducible} contributions.  These contributions
are integrands characterized by a subset of the original denominators
and an \emph{irreducible polynomial residue} for numerator.  The
multiple-cut conditions, which put these loop-momenta simultaneously
on-shell, can be viewed as \emph{projectors} isolating the
corresponding residue.  The residues have a universal
process-independent parametric form.  Some of the unknown
process-dependent coefficients appearing in this parametrization can
be identified with the ones which multiply the Master Integrals.

In Refs.~\cite{Zhang:2012ce,Mastrolia:2012an} the determination of the
residues at the multiple cuts has been formulated as a problem of
\emph{multivariate polynomial division}, and solved at any loop order
using algebraic-geometry techniques.  The most general parametric form
of a residue is indeed the most general remainder of a polynomial
division modulo the ideal generated by the corresponding subset of
denominators. The unknown coefficients of this parametrization can
then be computed, within the \emph{fit-on-the-cut approach}, by
evaluating the integrand on values of the loop momenta such that the
loop denominators which identify the residue are on-shell, as
traditionally done in the one-loop case.  In this paper, we also
present a two-loop example in $\N=4$ SYM.

The evaluation of the coefficients of the MI's of a one-loop
amplitudes can be improved by performing a Laurent expansion with
respect to the variables which are not fixed by the cut conditions, as
we showed in Ref.~\cite{Mastrolia:2012bu}.  The algorithm has been
implemented in the semi-numerical {\sc C++} library {\sc Ninja}, which
proved to be faster and numerically more stable than the original
integrand-reduction approach.  The library has been used for the
computation of NLO QCD corrections to Higgs boson production in
association with a top quark pair and a jet~\cite{vanDeurzen:2013xla}.

One can also perform the full integrand decomposition of any
multi-loop amplitude by means of purely algebraic operations, within
what we call the \emph{divide-and-conquer
  approach}~\cite{Mastrolia:2013kca}.  In this case, the residues are
generated by performing successive polynomial divisions between the
numerator and the denominators of a diagram.  Because of its wider
range of applicability we may consider the latter a more general
method for the integrand decomposition of loop integrals.

\section{Integrand reduction formula}
\label{sec:intredformula}
An arbitrary $\ell$-loop graph represents a $d$-dimensional integral of the form
\begin{align}
\int \, d^d q_1 \cdots  d^d q_\ell  \; \mathcal{I}_{i_1\cdots i_n}  \, ,\qquad \qquad
\mathcal{I}_{i_1\cdots i_n }
\equiv \frac{\mathcal{N}_{i_1 \cdots i_n}
 }{D_{i_1}\cdots D_{i_n}} \, ,
\label{Eq:integrand}
\end{align}
where $i_1,\ldots ,i_n$ are (not necessarily distinct) indices
labeling loop propagators.  The numerator $\mathcal{N}$ and the
denominators $D_i$ are polynomials in a set of coordinates $\z$, which
correspond to the components of the loop momenta with respect to some
basis.  Let $P[\z]$ be the ring of all polynomials in such
coordinates.  Every set of indices $\{i_1,\ldots ,i_n\}$ defines the
ideal
\begin{align}
 \mathcal{J}_{i_1i_2\cdots i_n}  \equiv {}&  \la D_{i_1}, \ldots ,D_{i_n} \ra \, 
=\,   \Bigg \{  \sum_{k=1}^n \,  h_k(\z)  \;
  D_{i_k}(\z) \; :  \; h_k(\z) \in P[\z]   \, 
\Bigg \} \, .
\label{Eq:JEJ}
\end{align}
The goal of the integrand reduction is finding a decomposition of the
form
\begin{equation*}
  \mathcal{I}_{i_1 \cdots i_n}
\equiv \frac{\mathcal{N}_{i_1 \cdots i_n}
 }{D_{i_1}\cdots D_{i_n}} \, =
 \sum_{k=0}^{n} \sum_{\{j_1\cdots j_k \}} \, \frac{\Delta_{j_1\cdots j_k}}{D_{j_1}\cdots D_{j_k}}
\label{Eq:integranddec}
\end{equation*}
where the \emph{residues} $\Delta_{j_1 \cdots j_k}$ are
\emph{irreducible} polynomials, i.e.\ polynomials which contain no
contribution belonging in the corresponding ideal $\mathcal{J}_{i_1
  \cdots i_n}$.

The numerator $\mathcal{N}$ of the integrand can be decomposed by
performing the \emph{multivariate polynomial division} modulo a
Gr\"obner basis $\GG{i_1\cdots i_n}$ of $\mathcal{J}_{i_1\cdots i_n}$
as
\begin{align}
\mathcal{N}_{i_1  \cdots i_n} ={} 
\Gamma_{i_1  \cdots i_n}
+{} \Delta_ {i_1  \cdots i_n} = \sum_{k=1}^n \mathcal{N}_{i_1 \cdots i_{k-1} i_{k+1} \cdots i_n} D_{i_k} +{} \Delta_ {i_1  \cdots i_n} 
\label{Eq:Enne}
\end{align}
in terms of a quotient $\Gamma_{i_1 \cdots i_n}$ and the remainder
$\Delta_ {i_1 \cdots i_n}$.  The properties of Gr\"obner bases ensure
that the remainder is irreducible, therefore it is identified with the
residue of the multiple cut $D_{i_1} = \cdots = D_{i_n} =0$, as
suggested by the notation.  The quotient $\Gamma$ belongs instead to
the ideal $\mathcal{J}$, hence in the last equality of
Eq.~\eqref{Eq:Enne} it has been written as combination of
denominators.  Substituting Eq.~(\ref{Eq:Enne}) in
Eq.~(\ref{Eq:integrand}), we obtain the recursive
formula~\cite{Mastrolia:2012an,Mastrolia:2013kca}
\begin{equation}
\mathcal{I}_{i_1 \cdots i_n}  =
  \sum_{k=1}^n \,  \mathcal{I}_{i_1 \cdots i_{k-1} i_{k+1} \cdots i_n} + {}
 \frac{\Delta_{i_1\cdots i_n}  }{ D_{i_1}\cdots D_{i_n} }.
 \label{Eq:Recurrence}
\end{equation}
Eq.~\eqref{Eq:Recurrence} expresses a given integrand in terms of an
irreducible residue sitting over its denominators and a sum of
integrands corresponding to sub-diagrams with fewer loop propagators.
Hence, the recursive application of this formula ultimately yields the
full decomposition of any integrand in terms of irreducible residues
and denominators, as in Eq.~\eqref{Eq:integranddec}.

The existence of such a recursive formula proves that the integrand
decomposition can be extended at any number of loops and the most
general parametrization of a residue $\Delta_{i_1\cdots i_n}$ can be
identified with the most general remainder of a polynomial division
modulo the Gr\"obner basis $\GG{i_1\cdots i_n}$.  Within the
\emph{fit-on-the-cut} approach the unknown coefficients of this
parametrization can be found by evaluating the numerator on the
solutions of the multiple cut $D_{i_1}=\cdots =D_{i_n}=0$.  This
method has been traditionally used at one loop, and recently applied
to higher-loop amplitudes as well
\cite{Mastrolia:2011pr,Badger:2013gxa,Mastrolia:2012wf,Badger:2012dv,Badger:2012dp}.

In Ref.~\cite{Mastrolia:2012an}, we applied the recursive formula in
Eq.~\eqref{Eq:Recurrence} to the most general one-loop integrand.
This allowed to easily derive the well know OPP decomposition for
dimensionally-regulated one-loop amplitudes
\cite{Ossola:2006us,Ellis:2007br}, as well as its higher-rank
generalization for effective and non-renormalizable theories
\cite{Mastrolia:2012bu} implemented in {\sc Xsamurai} 
(which extends the {\sc Samurai} library \cite{Mastrolia:2010nb}) and
recently used in the computation of NLO QCD corrections to Higgs boson
production plus two~\cite{vanDeurzen:2013rv} and three
jets~\cite{Cullen:2013saa} in gluon fusion, in the infinite top-mass
approximation.

\section{Integrand-Reduction via Laurent Expansion with {\sc Ninja}}
An improved integrand-reduction method for one-loop amplitudes was
presented in~\cite{Mastrolia:2012bu}, elaborating on the the
techniques proposed in~\cite{Forde:2007mi,Badger:2008cm}.  This method
allows to compute the coefficients of the Master Integrals, through
the systematic application of the Laurent series expansion on the
integrand, with respect to one of the free components of the loop
momenta which are not fixed by the on-shell conditions.

Within the original integrand reduction algorithm~\cite{Ossola:2007ax,
  Mastrolia:2008jb, Mastrolia:2010nb}, the determination of the
unknown coefficients requires to sample the numerator on a finite
subset of the on-shell solutions, subtract from the integrand all the
non-vanishing contributions coming from higher-point residues, and
solve the resulting linear system of equations.  Since in the
asymptotic limit both the integrand and the higher-point subtractions
exhibit the same polynomial behavior as the residue, with the
Laurent-expansion method one can instead identify the unknown
coefficients with the ones of the expansion of the integrand,
corrected by the contributions coming from higher-point residues.  In
other words, with this approach the system of equations for the
coefficients becomes diagonal and the subtractions of higher-point
contributions can be implemented as \emph{corrections at the
  coefficient level} which replace the subtractions at the integrand
level of the original algorithm.  The parametric form of this
corrections can be computed once and for all, in terms of a subset of
the higher-point coefficients.

This reduction algorithm has been implemented in the semi-numerical
{\sc C++} library {\sc Ninja}, which has been interfaced with the
package {\sc GoSam}
\cite{Cullen:2011ac,\comment{pierpaolo,}Gehrmann:2013sla\comment{,gionata,johannes}} for
automated one-loop computations.  Since the integrand of a loop
amplitude is a rational function, its semi-numerical Laurent expansion
has been implemented as a simplified \empty{polynomial division}
between the numerator and the denominators.

The input of the algorithm implemented in {\sc Ninja} is the numerator
cast in four different forms, which can be easily and very quickly
generated from the knowledge of the analytic dependence of the
integrand on the loop momentum (e.g.\ from the analytic expression of
the numerator generated by {\sc GoSam}).  The first form corresponds
to a simple evaluation of the numerator as a function of the loop
momentum, while the others return the leading terms of a parametric
Laurent expansion of the numerator.  {\sc Ninja} computes the
parametric solutions of each multiple cut, performs the Laurent
expansions via a simplified polynomial division between the
(expansions of the) numerator and the denominators, and implements the
subtractions at the coefficient level in order to get the unknown
coefficients.  These are then multiplied by the corresponding MI's.
{\sc Ninja} implements a wrapper of the {\sc OneLoop} library
\cite{vanHameren:2010cp,vanHameren:2009dr} which caches the values of
computed integrals and allows for constant time lookups from their
arguments.  The library can also be used for the reduction of
higher-rank integrands where the rank of a numerator can exceed the
number of denominators by one.  The simplified fit of the coefficients
and the subtractions at coefficient level make the algorithm
implemented in {\sc Ninja} significantly lighter, faster and more
stable than the original.

The first new phenomenological application of {\sc Ninja} has been the
computation of NLO QCD corrections to Higgs boson production in
association with a top quark pair and a jet~\cite{vanDeurzen:2013xla}.
The possibility of exploiting the improved stability of the new
algorithm has been especially important for the computation of the
corresponding six-point virtual amplitude, given the presence of two
mass scales as well as massive loop propagators which make traditional
integrand reduction algorithms numerically unstable.  Indeed, for the
highly non-trivial process under consideration, only a number of
phase-space points of the order of one per mill were detected as
unstable. All these points have been recovered using the tensorial
reduction provided by {\sc
  Golem95}~\cite{Binoth:2008uq,Cullen:2011kv}.

\section{Application to Two-Loop Scattering Amplitudes in $\mathcal{N}=4$ SYM}
The integrand reduction within the fit-on-the-cut approach has been
combined with the color-kinematic duality \cite{Bern:2008qj} in order
to construct the two-loop five-point amplitude for $\mathcal{N}$=4
super Yang-Mills (sYM) \cite{diplomuli}.

\begin{figure}
	\centering
		\includegraphics[width=0.5\textwidth]{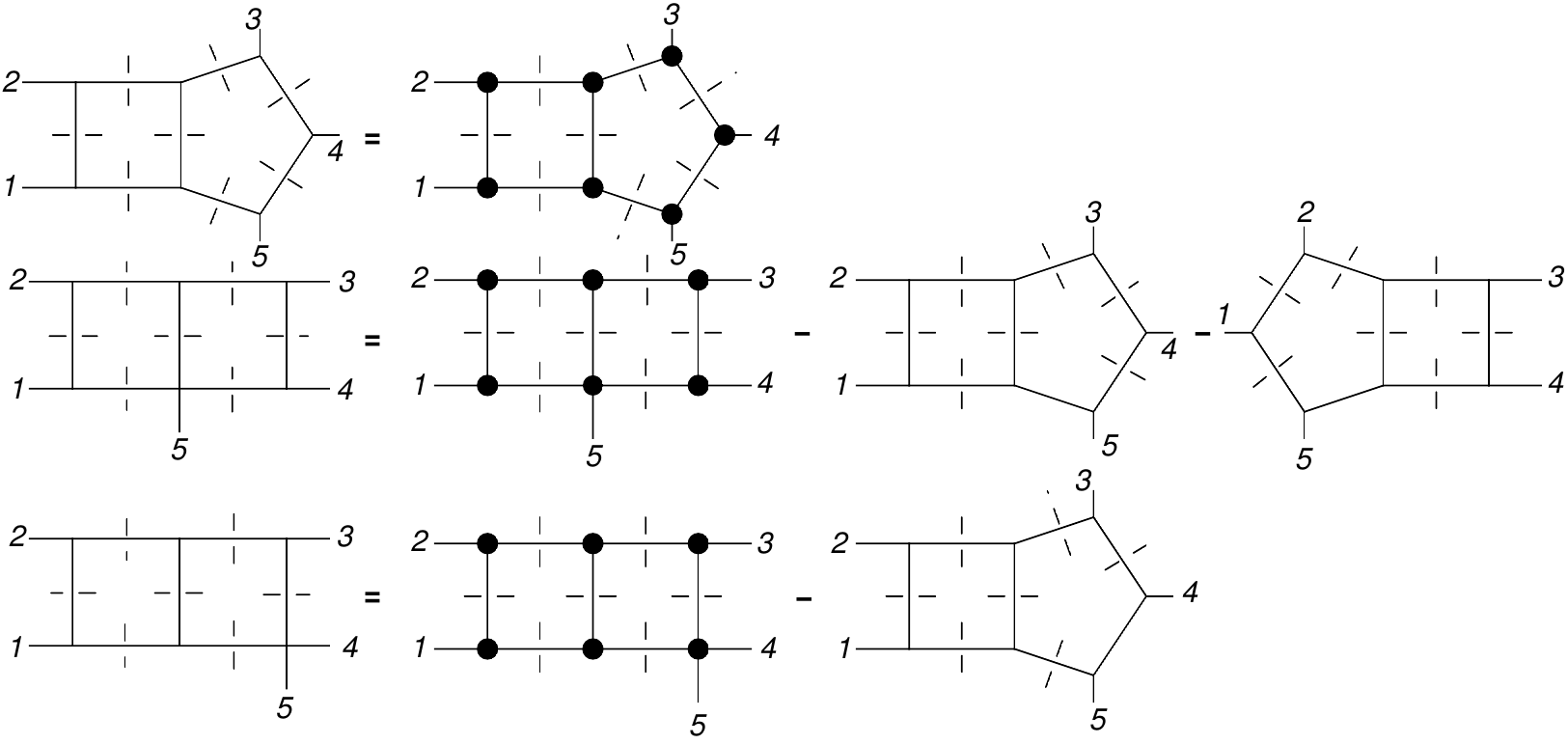}
        \caption{The eight- and seven-pole unitarity cuts of the
          pentabox graph, which are used to determine the
          corresponding residues.}
	\label{fig:unicut}
\end{figure}

\begin{figure}
	\centering
		\includegraphics[width=0.3\textwidth]{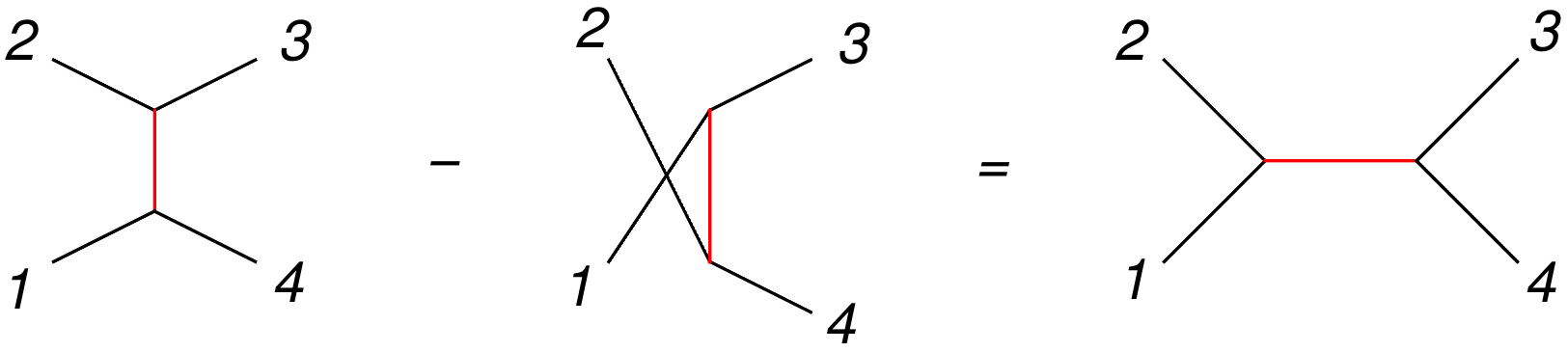}
	\caption{Graph representation of the Jacobi Identity at four points}
	\label{fig:jacobicolor}
\end{figure}

The inputs for the integrand decomposition are the products of the
trees to be sewn along the multiple cuts and the parametric forms of
the residues.  The former can be computed by adopting the
super-amplitude formalism \cite{Drummond:2008bq}, while the latter
were classified in \cite{Mastrolia:2012wf}.

The reduction of this amplitude begins from the eight-pole cuts, which
are maximum cuts \cite{Mastrolia:2012an}, and terminates after
determining the residues at the seven-pole cuts.  The absence of lower
cuts is compatible with the property that $\N=4$ sYM integrands are
linear in the loop-momenta.  Representative steps of the reduction are
given in Fig.~\ref{fig:unicut}.
 
\begin{figure}
	\centering
		\includegraphics[width=0.6\textwidth]{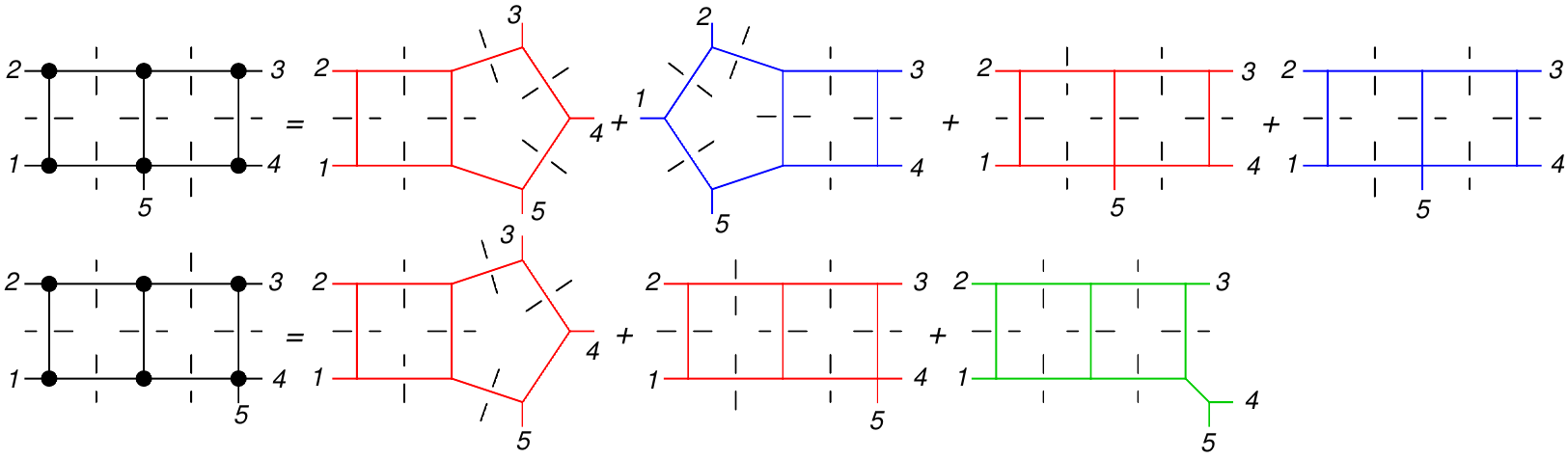}
	\caption{Each seven-pole unitarity cut is split into two contributions as indicated by the colors. The green diagram represents the new seven-denominator topology arising from the BCJ equations.}
	\label{fig:bcjpic1}
\end{figure}

\begin{figure}
	\centering
		\includegraphics[width=0.6\textwidth]{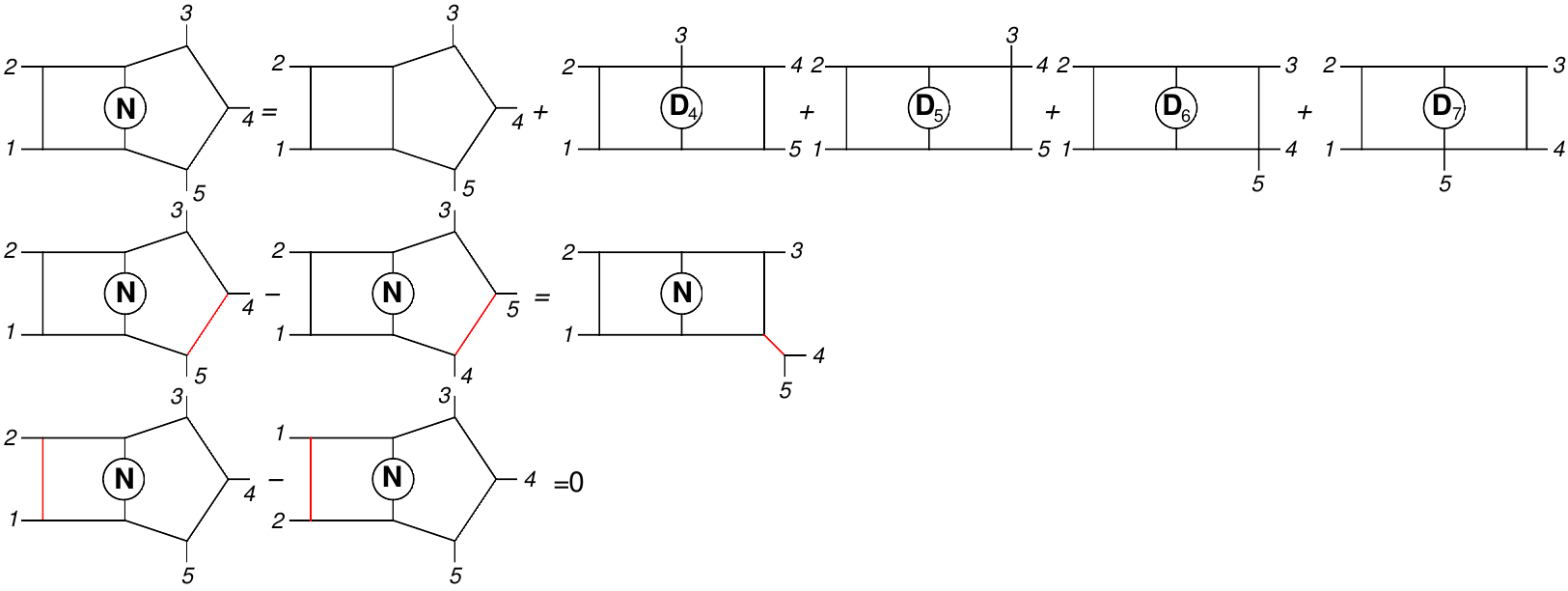}
        \caption{The first line shows the decomposition of the
          numerator $N$ of the planar topology, in terms of residues
          of eight-pole cuts and seven-pole cuts multiplied by the
          corresponding denominators $D_i$.  The last two lines show
          the BCJ equations which are used to disentangle the
          seven-pole unitarity cuts.}
	\label{fig:bcjpic2}
\end{figure}

Once the reduction is completed, one can equivalently construct a
numerator function for the parent eight-denominator topologies, which
captures the whole structure of the scattering amplitude. These
numerators can be rearranged in a color-kinematic dual form by
imposing additional constraints, referred to as BCJ equations, derived
from the kinematic equivalent of the Jacobi Identity portrayed in
Fig.~\ref{fig:jacobicolor}.  BCJ identities, beyond one-loop, imply a
relation between the integrands of planar and non-planar topologies.
The color-kinematic dual numerator for the eight-denominator planar
diagram is represented in the first line of Fig.\ref{fig:bcjpic2}.
The key equations for the determination of the new numerators of the
planar diagrams are depicted in Fig.~\ref{fig:bcjpic1}, where one may
notice the rising of a seven-denominator diagram, whose identification
was not needed in the unitarity decomposition.
In order to disentangle the contributions to the seven-pole cuts we
use the BCJ equations which only involve the planar topologies
displayed in the last two lines of Fig.~\ref{fig:bcjpic2}. The
obtained results are in agreement with \cite{Carrasco:2011mn}.

\section{Divide-and-conquer approach}
The direct application of the integrand reduction formula of
Eq.~\eqref{Eq:Recurrence} on the numerator of an $l$-loop graph allows
to perform the integrand decomposition algebraically by successive
polynomial divisions, within what we call the
\emph{divide-and-conquer} approach~\cite{Mastrolia:2013kca}.  At each
step, the remainders of the divisions are identified with the residues
of the corresponding set of denominators, while the quotients become
the numerators of the lower-point integrands appearing on the r.h.s.\
of the formula, allowing thus to iterate the procedure.  In this way,
the decomposition of any integrand is obtained analytically, with a
finite number of algebraic operations, without requiring the knowledge
of the varieties of solutions of the multiple cuts, nor the one of the
parametric form of the residues.

\begin{figure}[htb]
\centerline{%
\includegraphics[width=\textwidth]{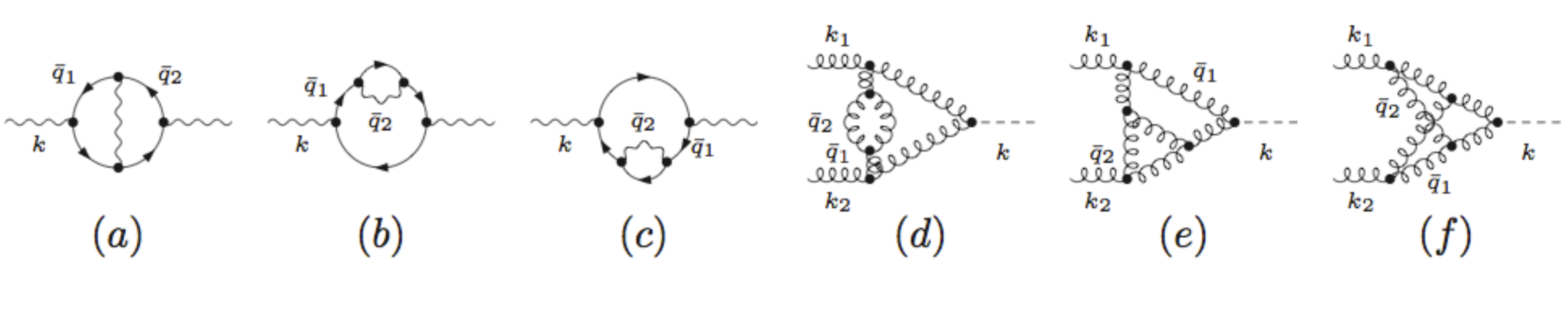}}
\caption{Examples of diagrams reduced using the divide-and-conquer approach.}
\label{Fig:diags}
\end{figure}

This algorithm has been implemented in a {\sc Python} package, which
can perform the decomposition of any numerator using {\sc
  Form}~\cite{Vermaseren:2000nd} and {\sc Macaulay2}~\cite{Macaulay2}
for the algebraic operations.  It has been applied to the examples
depicted in Fig.~\ref{Fig:diags}.  Despite their simplicity, these
show the broadness of applicability of the method which is not
affected by the presence of massive propagators, non planar diagrams,
higher powers of loop denominators or higher-rank contributions in the
numerator.

\section{Conclusions}
We described a coherent framework for the decomposition of Feynman
integrals, which can be applied at any loop order, regardless of the
complexity of the integrand, the number of external legs or the
presence of higher powers of loop denominators.  This framework allows
to easily derive well known results at one-loop order and extend them
to higher loops.

In the one-loop case, we showed how the knowledge of the analytic
structure of the integrands on the multiple cuts, and in particular
their asymptotic behavior on the on-shell solutions, can be used to
improve the analytic and semi-analytic reduction with the Laurent
expansion method.  Its implementation in the {\sc C++} library {\sc
  Ninja} provided a considerable gain in the speed and in the
stability of the reduction.

At higher loops we presented the application of the
\emph{fit-on-the-cut approach} to 5-point amplitudes in
$\mathcal{N}$=4 super Yang-Mills (sYM), with a unitarity-based
construction of the integrands on the multiple cuts.  We also
described the application of the \emph{divide-and-conquer} approach,
which allows to perform the full decomposition with purely algebraic
operations,
without requiring the knowledge of the algebraic variety
defined by the on-shell solutions.  We applied it to simple examples,
some of which cannot be addressed with other unitarity-based and
integrand-reduction methods, due to the presence of higher powers of
loop denominators in the integrands.  Since it is based on the same
principles used to constructively prove the existence of the integrand
decomposition at all loops, the \emph{divide-and-conquer} approach
doesn't have the limitations of other methods and can be considered a
more general integrand reduction algorithm.

\section*{Acknowledgments}
The work of H.v.D., G.L., P.M., T.P.\ and U.S.\ was supported by the
Alexander von Humboldt Foundation, in the framework of the Sofja
Kovalevskaja Award Project ``Advanced Mathematical Methods for
Particle Physics'', endowed by the German Federal Ministry of
Education and Research.  G.O.\ was supported in part by the National
Science Foundation under Grant PHY-1068550.

\end{document}